# Controlling the oxidation of magnetic and electrically conductive solid-solution iron-rhodium nanoparticles synthesized by Laser Ablation in Liquids

Ruksan Nadarajah[1], Shabbir Tahir[1], Joachim Landers[2], David Koch[3], Anna S. Semisalova[2], Jonas Wiemeler[2], Ayman El-Zoka[4], Se-Ho Kim[4], Detlef Utzat[2], Rolf Möller[2], Baptiste Gault[4,5], Heiko Wende[2], Michael Farle[2], Bilal Gökce[1,*]

[1] University of Duisburg-Essen, Technical Chemistry I and Center for Nanointegration Duisburg-Essen (CENIDE), Universitaetsstr. 7, 45141 Essen, Germany
[2] University of Duisburg-Essen, Faculty of Physics and Center for Nanointegration Duisburg-Essen (CENIDE), Lotharstr. 1, 47057 Duisburg, Germany
[3] Institute of Materials Science, University of Technology, Darmstadt 64287, Alarich-Weiss-Strasse 2, Germany
[4] Max-Planck-Institut für Eisenforschung GmbH, Max-Planck-Strasse 1, 40237 Düsseldorf, Germany
[5] Department of Materials, Royal School of Mines, Imperial College London, London, SW7 2AZ, UK

Correspondence: bilal.goekce@uni-due.de; Tel.: +49201/183-3146

**Abstract:** This study focuses on the synthesis of FeRh nanoparticles via pulsed laser ablation in liquid and on controlling the oxidation of the synthesized nanoparticles. Formation of monomodal γ-FeRh nanoparticles was confirmed by transmission electron microscopy (TEM) and their composition confirmed by atom probe tomography (APT). On these particles, three major contributors to oxidation were analysed: 1) dissolved oxygen in the organic solvents, 2) the bound oxygen in the solvent and 3) oxygen in the atmosphere above the solvent. The decrease of oxidation for optimized ablation conditions was confirmed through energy-dispersive X-ray (EDX) and Mössbauer spectroscopy. Furthermore, the time dependence of oxidation was monitored for dried FeRh nanoparticles powders using ferromagnetic resonance spectroscopy (FMR). By magnetophoretic separation, B2-FeRh nanoparticles could be extracted from the solution and characteristic differences of nanostrand formation between γ-FeRh and B2-FeRh nanoparticles were observed.

## 1. Introduction

Magnetic alloy nanoparticles offer a broad range of applications in catalysis [1-3], biomedicine [4-6], and information technology [7, 8]. Many applications require a colloidal form of the particles in water, organic solvents, or polymer solutions for nanoparticle-polymer composites [9, 10]. These nanoparticles can be synthesised by various bottom-up and top-down approaches such as chemical precipitation [11, 12], solvothermal [13, 14], solution-phase [15, 16], and ball milling [17, 18]. Laser ablation in liquids (LAL) is a top-down approach, which needs no stabilizers or precursors to produce colloidal nanoparticles. [7] It has been proven to be a scalable and versatile process [19, 20] due to the free choice of the target material and solvent [19]. Often LAL leads to the formation of oxides or metal complexes during ablation of metals. Even though magnetic metal oxide nanoparticles are omnipresent in science and technology, also metallic magnetic nanoparticles are needed for many applications for use in magnetic composites [21] or catalysis [22]. Hence, in this study we want to explore ways of minimizing the oxidation level of magnetic metal nanoparticles.

A challenging material in terms of oxidation is $Fe_{50}Rh_{50}$, which forms a body-centered cubic (bcc) structure known for its first order antiferro-ferromagnetic (AFM-FM) transition at T = 370 K [23]. This transition takes place only near the equimolar ratio of FeRh with a lattice expansion of about 1%, an increase in entropy and net magnetic moment, and a decrease in electrical resistivity. A further magnetic phase transition occurs at the Curie temperature $T_C$ = 650 K from the ferromagnetic to the paramagnetic phase. Since the discovery of the magnetocaloric effect at the first-order phase transition, the $Fe_{50}Rh_{50}$ system has extensively been investigated by different groups for heat-assisted magnetic recording [24, 25], magnetocaloric cooling systems [26, 27], (bio-)hyperthermia [28, 29], and catalysis [30, 31]. The phase transition temperature is known to be very sensitive to changes near the equiatomic composition [32].

Reports on the synthesis of FeRh nanoparticles can be found in the literature. Typically wet chemical synthesis routes are used to produce nanoparticles for magnetic and catalytic applications [30, 33, 34]. In the field of catalysis, FeRh nanoparticles can reduce olefins and nitroarenes and also promote the conversion of syngas to $C^{2+}$ oxygenates[1, 30]. For many applications such as magnetic refrigeration, a composition near the equimolar composition is important e.g. to attain an AFM-FM transition at higher temperatures. The composition control of nanoparticles, however, is difficult to achieve using wet chemical approaches. In all cases using wet-chemistry, the synthesis of FeRh leads to the formation of core-shell NPs [34] or an alloy with inhomogeneous composition [35]. In addition, all of the synthesis methods lead to the formation of FeRh nanoparticles with particle sizes smaller than 20 nm. And due to the high surface to volume ratio, these smaller nanoparticles are more prone to oxidation.[33, 35, 36]

Since the properties of these alloys are sensitive to the change of composition, the control of oxidation becomes very important. Oxides have often different properties in terms of magnetism, catalytic activity, and electric/thermal conductivity than their metal counterparts. In the laser ablation process in general, the formed particles and the target are in an environment containing oxygen, which will inevitably lead to oxidation. Here, we report on the control and minimization of the oxidation of FeRh alloy nanoparticles by reducing possible oxidation sources during the laser ablation process in liquid.

## 2. Materials and Methods

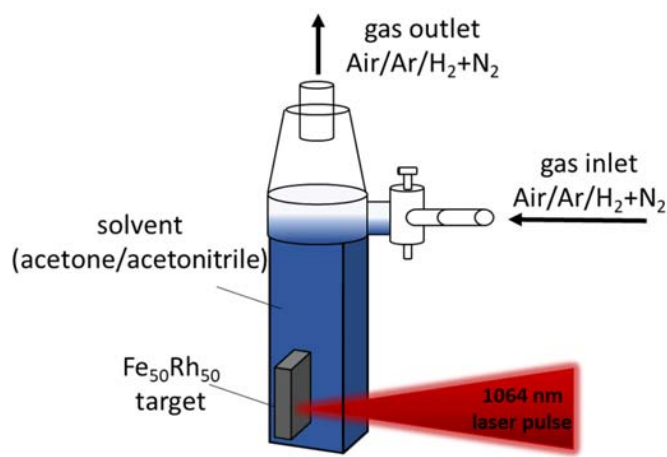

**Figure 1.** Schematic illustration of the "Schlenk" ablation chamber. The figure displays the laser setup and the different configurations with respect to solvents (acetone or acetonitrile) and gas flow. The gas inlet allows the manipulation of the atmosphere, with air, argon and a mixture of nitrogen and hydrogen. A laser with a pulse duration of 10 ps, repetition rate of 100 kHz and wavelength of 1064 nm was used.

In figure 1 the schematic illustration of the experimental setup is shown. The laser pulse interacts with the target and creates a plasma plume, which consists of ionized and atomized species. During the plasma decay, the energy is transferred to the surrounding liquid which will evaporate and form a cavitation bubble, in which the nanoparticles are formed. The ablation is performed in a Schlenk-like ablation chamber, which allows the ablation in different atmospheres. Besides air, argon as an inert gas, and a mixture of 95% nitrogen and 5% hydrogen as a reducing gas was used to minimize the oxidation of the synthesized nanoparticles by preventing the exchange of oxygen from the atmosphere into the liquid. The chamber consists of two parts. The top part was built with a three-way valve for gas-flow control. The lower part, where the ablation takes place, is made of a quartz glass cuvette (path length 0.5 cm) which is transparent for the laser wavelength of 1064 nm. More information about this setup can be found elsewhere [37].

Here, LAL was performed in an organic solvent (acetone, acetonitrile) using a picosecond pulsed Nd-YAG laser (Ekspla, Atlantic Series, 10 ps, 100 kHz, 80 µJ, 1064 nm). The laser beam was directed into a laser scanner and focused by an f-theta lens (focal length of 100.1 mm). The laser beam has a Gaussian profile with a laser fluence of 3.5 J/cm$^2$. A scanning speed of 6 m/s was chosen, to spatially bypass the laser-induced cavitation bubbles and neglect the influence of the scanning speed [38]. The particle morphology, composition, and size distribution was analysed by transmission electron microscopy (TEM) (JEOL 2200FS) and atom probe tomography (APT) (Cameca 5076XS). Phase analysis was conducted by X-ray diffraction (XRD). The XRD measurements were performed using Mo K$\alpha$ radiation in transmission geometry on a custom-built setup with a Mythen2 R 1K detector (Dectris Ltd). The sample was mixed with NIST 640d standard silicon powder for correcting geometric errors and glued on a graphite foil. Energy-dispersive X-ray spectroscopy (EDX) in the dried form and Mössbauer spectroscopy on colloidal samples were conducted to test the oxidation level and to characterize the particles' structural and magnetic properties. Mössbauer spectra were taken in transmission geometry and constant acceleration mode at 4.3 K and ca. 80 K using a l-He bath cryostat. The ferromagnetic resonance (FMR) experiments were performed using a Bruker Elexsys E500 EPR spectrometer in the X-band (9.5 GHz) via conventional cavity-based FMR detection. The observed signal is proportional to the B-field derivative of the imaginary part of magnetic susceptibility, $\partial\chi''/\partial B$. Time-dependent FMR measurements were performed to study the effect of oxidation on the resonance field and lineshape of the FeRh nanoparticles. Initially, a small amount of the solution (0.013 mL) containing the nanoparticles in acetone was filled into a quartz tube (2 mm ID), and the first measurement was taken.

In order to fabricate APT specimens, freestanding nanoparticles first need to be encapsulated. The as-synthesized FeRh nanoparticles were originally suspended in acetone. To remove the acetone, the suspension was left to dry in a nitrogen glovebox. The collected nanoparticles were co-electrodeposited with Ni according to the protocol detailed in ref.[39]. The dried FeRh NP powder was dispersed in a Ni ion electrolyte followed by pouring the solution into a vertical electro-cell where a counter electrode (Pt mesh) positioned on top and a working electrode (Cu foil) placed on bottom. Then, a low constant current (-19 mA) was applied to the electrodes to co-electrodeposit Ni and NPs on the working electrode. After the deposition, the composite film was kept in a vacuum desiccator.

The electroplated composite film was loaded into the focused ion beam (FIB) (Helios 600) and needle-shaped specimens were prepared following the in-situ lift-out protocol outlined in ref. [40]. APT measurements were done using a UV-laser assisted LEAP™ 5000 XS system (CAMECA Instruments Inc.) at an analysis stage temperature of 40 K. The laser repetition rate was set to 200 kHz, and the laser pulse energy was 60 pJ. Data was acquired with a detection rate of 5 ions for 1000 pulses on average. Multiple datasets containing over 15 million ions were collected. Data analyses were performed using the commercial software, IVAS 3.8.4.

Furthermore, nanostrands were formed by producing a solution of 5 wt% poly(methylmethacrylate) (PMMA) with 0.2 wt% FeRh NPs in acetone. The solution was dried on a glass substrate in an external magnetic field, with a flux density of 150 mT. The dried sample with the formed nanostrands were imaged using an optical microscope (CX 40, Olympus). More information on the setup and formation mechanism can be found elsewhere [9]. The conductivity of the nanostrands was measured by using thin-film gold interdigitated electrodes (Micrux).

The ablation target was custom-made by the "Research Institute for Noble Metals and Metal Chemistry", Schwaebisch Gmuend, Germany and has an equimolar composition (±1.5%) according to X-ray fluorescence (XRF) and energy-dispersive X-ray (EDX) measurements.

## 3. Results and discussion

To minimize oxidation of the alloy nanoparticles, 3 major contributors to oxidation were analysed. 1) Dissolved oxygen in the organic solvents, 2) bound oxygen in the solvent due to the dissociation of the solvent during the LAL process [41] and 3) oxygen in the air. To study the influence of bound oxygen, acetone and acetonitrile were used as solvents. Both are similar organic solvents with the difference that acetone has bound oxygen while acetonitrile contains nitrogen instead. Both solvents were successfully used in the literature to synthesize elemental metal nanoparticles ([37],[42]).

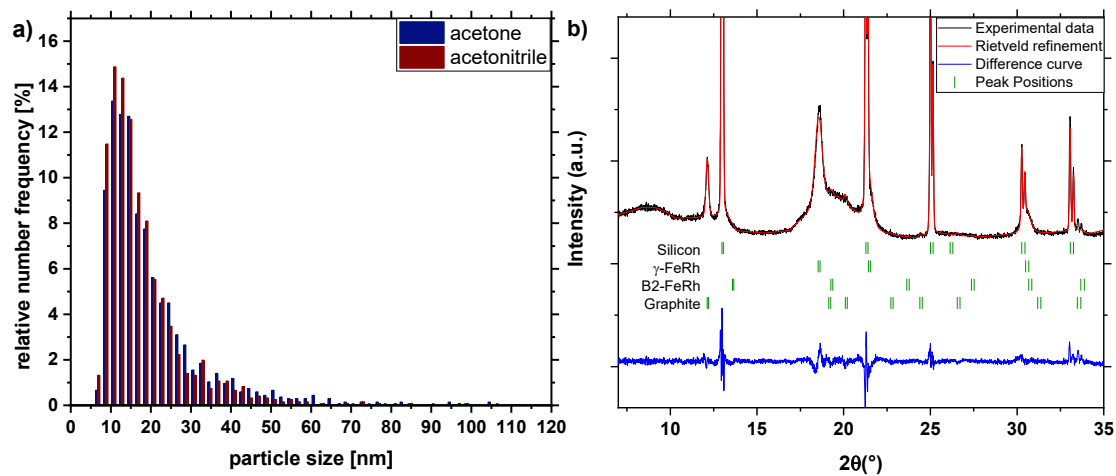

**Figure 2.** a) particle size distribution of synthesized particles via PLAL in acetone and acetonitrile and b) XRD-pattern of $Fe_{50}Rh_{50}$ nanoparticles obtained by laser ablation in acetone and acetonitrile. The known XRD peaks of $\gamma$—FeRh, graphite and B2-FeRh and silicon are indicated.

Figure 2 shows the number-weighted size distribution of $Fe_{50}Rh_{50}$ nanoparticles synthesized in acetone and acetonitrile under argon atmosphere. The size distribution can be fitted using a lognormal function, which leads to a mean particle size of 15.24 ± 0.39 nm (acetone, red) and 14.46 ± 0.37 nm (acetonitrile, blue), respectively. With a polydispersity index (PDI) lower than 0.3, both particle size distributions can be considered as monomodal. Nanoparticle productivities of 43.6 mg/h and 41.2 mg/h were reached in acetone and acetonitrile, respectively. XRD measurements of the particles in acetone with argon atmosphere show that there is almost no visible B2 phase, which corresponds to the $\alpha'$-FeRh phase (bcc), which is antiferromagnetic at room temperature, as we would expect from bulk FeRh near the equimolar ratio. Instead, almost all the particles are in the $\gamma$-phase, which was also found by Jia Z. et al. for FeRh nanoparticles of 4-20 nm size synthesized by a polyol coreduction process [33]. Rietveld refinement via the "Fullprof" software [43] shows that almost all FeRh nanoparticles are in the $\gamma$-phase, with a lattice parameter of 3.81 Å. The formation of the $\gamma$-phase is also supported by density functional theory calculations [44] which showed that due to the similar energy states of AFM/BCC FeRh and metastable FCC FeRh epitaxial strain can easily transform the BCC to the FCC structure.

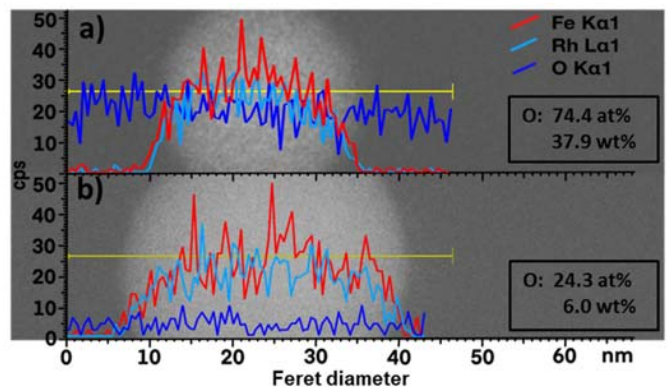

**Figure 3.** EDX line-scan of single $Fe_{50}Rh_{50}$ nanoparticles obtained by laser ablation in acetone a) with the use of mole sieve, and b) without the use of a mole sieve.

Figure 3 a) shows the EDX line scan for a representative particle which was synthesized in acetone under air atmosphere. The content of oxygen was found to be as high as 74.4 at% (37.9 wt%). Furthermore, an oxygen signal on spots without particles can be found. It can be assumed, that an unbound oxide layer originates from the drying step on the TEM grids. Water is soluble in air, and at 25°C, 1 m³ of air can hold up to 22 g of water at 100% humidity. As water is miscible in acetone and is one of the main suppliers of oxygen, a mole sieve with a pore size of 3 Å was used to remove the water content in the acetone, which could have been transferred from the air to the solvent. A reduction of the detected oxygen from 74.4 at% (37.9 wt%) to 24.3 at% (6.0 wt%) can be observed in figure 3. As water was found to be a major contributor for the oxidation, further experiments were performed with a mole sieve purified solvent.

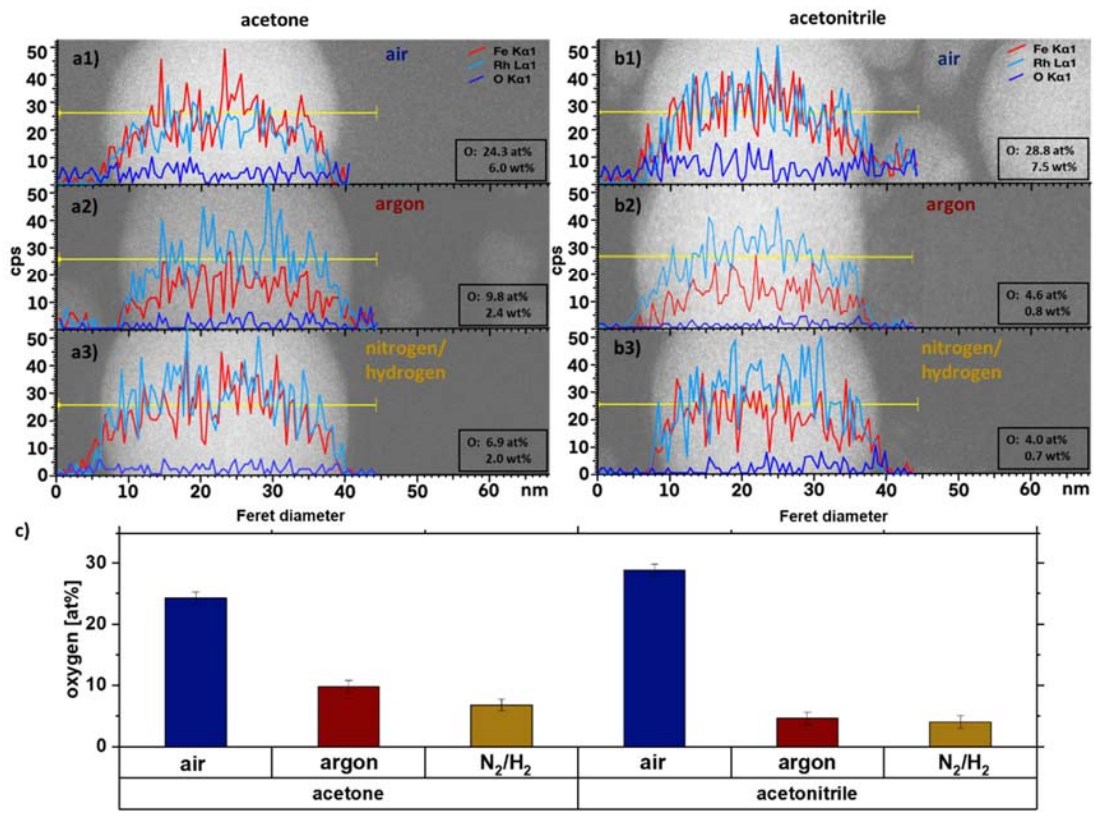

**Figure 4.** EDX line-scan of $Fe_{50}Rh_{50}$ nanoparticles synthesized in acetone under a1) air, a2) argon, a3) $N_2/H_2$ mixture atmosphere and in acetonitrile under b1) air, b2) argon, b3) $N_2/H_2$ mixture atmosphere. Panel c) depicts the extracted oxygen fraction for the different configurations.

The degree of oxidation of the synthesized iron-rhodium alloy nanoparticles was further examined in acetone and acetonitrile. As the surrounding atmosphere is also a potential oxygen supplier for oxidation, the ablation was conducted in 3 different atmospheres. As an atmosphere, besides air as a reference, argon was chosen as an inert gas. Further, the process was tested with a mixture of nitrogen and hydrogen as a reduction gas, which will eventually reduce the formed oxidation. Figure 4 summarizes EDX line scans for individual FeRh nanoparticles obtained for all configurations presented in the Table 1. The particle size can have an impact on the oxidation, wherefore particles of similar size of ~30nm were chosen to have comparable results. Furthermore, too small particles will result in a poor signal-to-noise ratio. As smaller particles have a higher surface-to-volume ratio and the oxide level of the smaller particles can therefore be much higher in comparison, EDX is used as a screening method and selected particles are further analysed by atom probe tomography (APT) and Mössbauer spectroscopy.

*Table 1 Configuration of solvent and atmosphere during PLAL.*

| Configuration | Solvent | Atmosphere |
|---|---|---|
| a1 | acetone | air |
| a2 | acetone | argon |
| a3 | acetone | $N_2/H_2$ |
| b1 | acetonitrile | air |
| b2 | acetonitrile | argon |
| b3 | acetonitrile | $N_2/H_2$ |

As seen in figure 4a, the EDX-line scan of the particles which were ablated in acetone under the air atmosphere showed an oxide level of 24.3 at%. When argon or the $N_2/H_2$ mixture was used as an atmosphere, the detected oxygen was reduced to 9.8 at% and 6.9 at%, respectively. M. Quaranta et al calculated that the solubility of oxygen in acetone and acetonitrile is 2.55 ± 0.19 mM and 2.60 ± 0.15 mM, respectively [44]. This means that the proportion of oxygen, which promotes oxidation in the solvent, is the same for both solvents. As the amount of solved oxygen in the solution is same, the change of oxidation would be solely due to bound oxygen in the acetone molecule. An effect of the bound oxygen could not be observed.

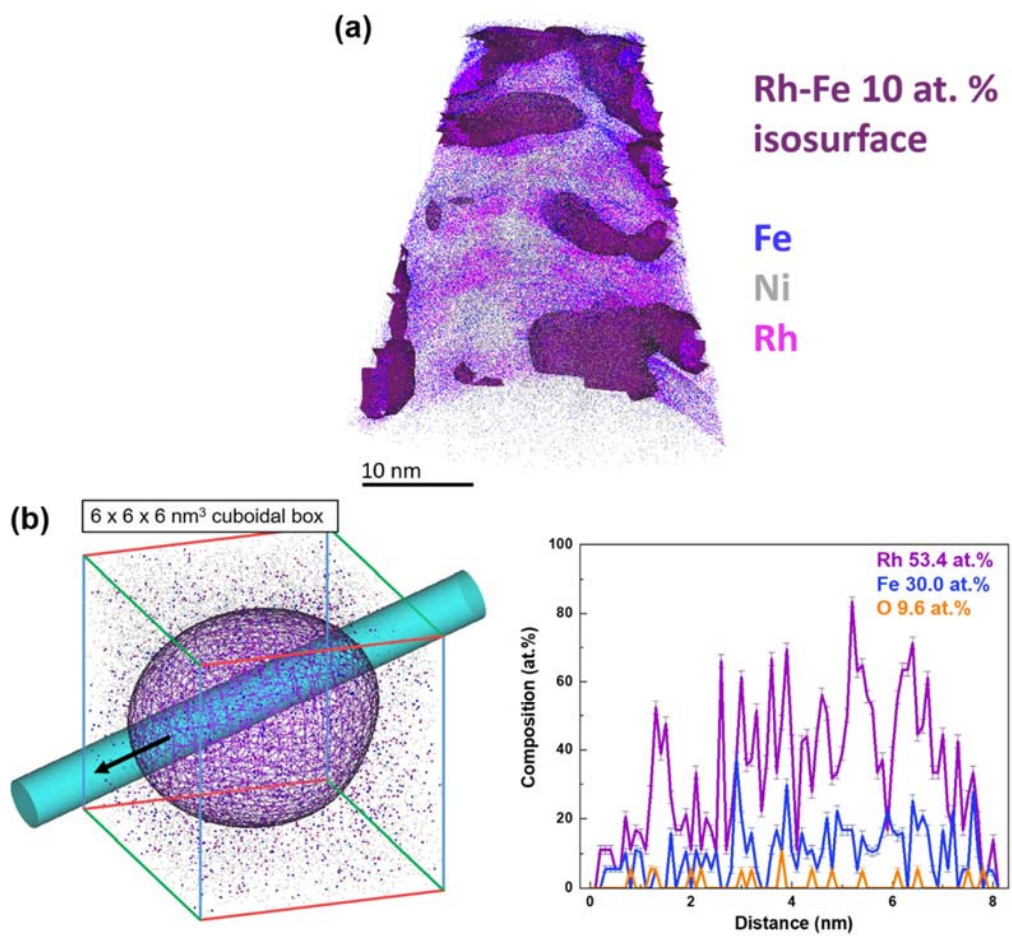

**Figure 5.** Atom probe tomography of FeRh nanoparticles synthesized in acetone and argon atmosphere. (a) 3D atom map of FeRh nanoparticles in Ni deposit. Blue, purple, and grey dots mark the reconstructed positions of Fe, Rh, and Ni, respectively. Detected FeRh nanoparticles are highlighted by 10 at. % FeRh iso-composition surfaces. (b) Extracted RhFe nanoparticle (25 at.% of FeRh iso-composition surface) and line scan of selected FeRh nanoparticle.

For more advanced characterization of the elemental distribution within the nanoparticles, we performed APT analysis on FeRh nanoparticles after electrochemical embedding in Ni (see Materials and Methods). Figure 5a shows a 3D atom map of Fe-Rh nanoparticles with representative sizes and respective chemical profiles in Ni deposit. The particle surface is highlighted using a set of iso-composition surfaces encompassing regions containing more than 10 at% (Fe+Rh). Interestingly the APT analysis of FeRh nanoparticles shows the homogenous distribution of Fe and Rh across the nanoparticles with different sizes (figure 5), which is in contrast to known literatures, where only core-shell particles were reported [34-36, 45]. The acquired particle in Figure 5b has a high composition of Rh while other shows the opposite with a slightly higher Fe composition (see **Supplementary Figure SX**). Moreover, in 1D compositional profile along FeRh nanoparticle, the O content of 9.6 at.% is detected, which could be originated not only from nanoparticle synthesis but also APT sample preparation for which aqueous electrolyte is used. Nevertheless, the levels of O detected inside the nanoparticle suggest the presence of oxygen inside the nanoparticles.

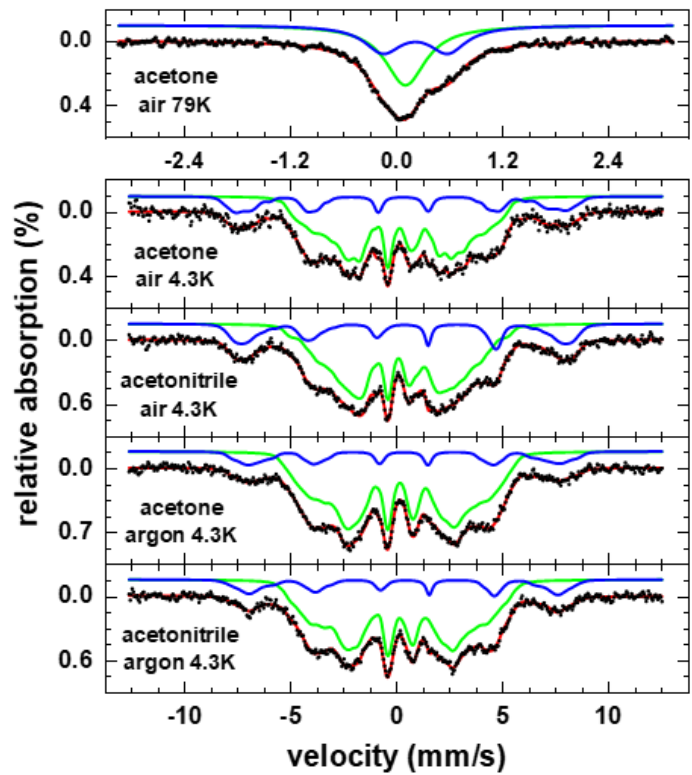

**Figure 6.** Mössbauer spectra of laser ablated FeRh nanoparticles in solution prepared under air/argon atmosphere in acetone or acetonitrile, respectively. Spectra were recorded at 4.3 K; subspectra can be assigned to metallic γ-phase FeRh (green) and an $Fe^{3+}$-oxide phase (blue). A spectrum obtained at 79 K is shown for comparison on top.

In figure 4, EDX was used as a screening method. As Mössbauer spectroscopy has the ability to directly distinguish metal and oxide constituents, it is used here to verify the EDX findings. To compare the results of both techniques, it also has to be considered that particles were measured in the dried state in EDX and in solution via Mössbauer spectroscopy and that the latter provides an integral signal averaged over the whole sample, unlike single particle analysis as done in EDX or APT. To allow measurements of the particles in solution, Mössbauer spectra as shown in figure 6 were recorded using a liquid-tight screw-mountable sample holder at 4.3 K and 79 K for samples prepared in acetone and acetonitrile in atmospheres of air and argon, respectively.

At 4.3 K we observe two magnetically ordered sextets, which can be assigned to metallic FeRh (green) and presumably a $Fe^{3+}$-oxide state based on their different average hyperfine magnetic fields of ca. 21 and 45.5 T and isomer shifts of ca. 0.29 and 0.45 mm/s relative to α-Fe at room temperature, respectively. Instead of narrow absorption lines, both subspectra are distinctly broadened, wherefore they were reproduced via distributions of hyperfine magnetic fields. For the oxide phase, this broadening could potentially be explained by being present as an amorphous oxide surface layer, hindering an in-detail analysis of this phase's spectral properties. For the metallic constituent, on the other hand, the observed distribution is in agreement with previous reports on antiferromagnetic γ-phase FeRh, although average hyperfine magnetic fields are somewhat higher here in comparison, which could be connected e.g. to minor variations in local stoichiometry or strain effects in the nanoparticles[46, 47]. The presence of γ-FeRh in particles studied here is also supported by the absence of magnetic ordering in the reference spectrum recorded at 79 K, consistent with the low ordering temperature of γ-FeRh[47], although in theory, superparamagnetic relaxation of smaller FeRh nanoparticles could also contribute to the doublet present at this temperature. The higher superposition of subspectra at 79 K prevents a precise estimation of metallic and oxide fraction, while at 4.3 K similar oxide fractions of ca. 16 % and 21 % are found for the sample prepared in argon and air, respectively (see also table 2), relatively independent of the used solvent. Assuming identical Lamb-Mössbauer factors for both components at 4.3 K, relative spectral areas as given here correspond in good approximation to the fraction of Fe-atoms in the respective phase (Table 2), in general yielding similar trends as EDX-analysis in terms of nanoparticle oxidation.

Table 2 Oxide fractions extracted from the Mössbauer spectra (fraction of Fe-atoms in the oxide phase).

|  | acetone | acetonitrile |
|---|---|---|
| **air** | 20.3 ± 1.7 at% | 21.5 ± 1.9 at% |
| **argon** | 15.7 ± 1.6 at% | 15.9 ± 1.4 at% |

Ablation in acetone and acetonitrile shows the same amount of oxide and corresponds to the EDX-measurements in figure 4., while the influence of the bound oxygen of the acetone molecule cannot be seen in the Mössbauer spectra. Furthermore, the ablation in air shows 20-22 at% of oxidation, while the ablation in the argon atmosphere results in an oxide content of 15-16 at%. Although the trend is the same as in the EDX line scans, it is difficult to compare absolute numbers determined by both methods: EDX provides atomic fraction of O-atoms, whereas Mössbauer spectroscopy provides the fraction of Fe-atoms, which are part of the oxide phase. Smaller particles, which cannot be measured with EDX, can have a higher degree of oxidation, and would also contribute to the oxide fraction extracted from Mössbauer spectra.

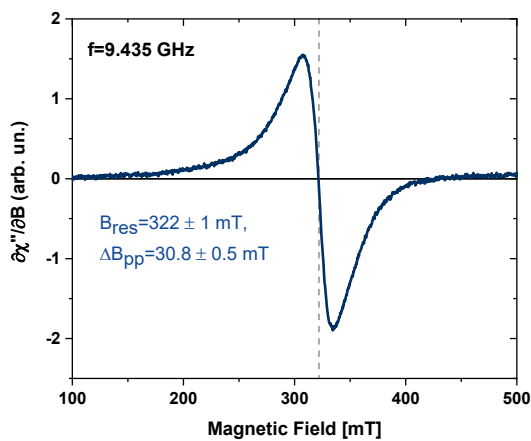

**Figure 7.** The ferromagnetic resonance spectrum measured at 9.435 GHz of a FeRh nanoparticles solution in acetone at room temperature. $B_{res}$ is the resonance field (dotted line), $\Delta B_{pp}$ is the peak-to-peak linewidth.

To probe the magnetization of the FeRh particles and to monitor the time dependent oxidation of the dried FeRh particles in air, ferromagnetic resonance measurements were conducted to probe the magnetization of the FeRh nanoparticles and to monitor its time dependence in air. Based on the EDX and Mössbauer spectroscopy results, we chose particles which were synthesized in acetone and argon atmosphere as this configuration was expected to lead to a minimum of oxidation.

In figure 7, a typical FMR spectrum is shown. The FeRh NPs in acetone was found to have an asymmetric lineshape, which is typical for an ensemble of ferromagnetic nanoparticles – the observed line is the sum of multiple FMR Lorentz-shape spectra from individual nanoparticles varying in size, shape, and orientation of the anisotropy axes [48, 49]. The FMR signal is proportional to the magnitude of the magnetization of the sample [50]. The resonance field is a measure of the internal magnetic field which consists of shape and magnetocrystalline anisotropy fields and includes possible dipolar coupling effects between particles. The peak-to-peak linewidth of the FMR spectrum of the nanoparticles ensemble reflects the distribution of anisotropy axes.

The total magnetic anisotropy of nanoparticles given by a combination of shape, magnetocrystalline, including surface anisotropy contributions, has been shown to be effectively describable by a uniaxial anisotropy field $B_A = \frac{2K_{eff}}{M_s}$, where $K_{eff}$ is the effective anisotropy constant and $M_s$ is the saturation magnetization. The resonance field value $B_{res}$=322±1 mT observed for the ensemble of FeRh nanoparticles in acetone (fig. 7) can be analyzed within the model of randomly distributed easy axes of magnetization[51]. In this model, the measured resonance field is shifted to lower values from the resonance field for zero anisotropy. Assuming the g-factor for $Fe_{50}Rh_{50}$ nanoparticles g=2.05±0.02 [45, 52] corresponding to an isotropic resonance field $B_0 = \frac{h\nu}{g\mu_B} = 330 \pm 3$ $mT$, one obtains a small anisotropy field $B_A$=47±10 mT at room temperature.

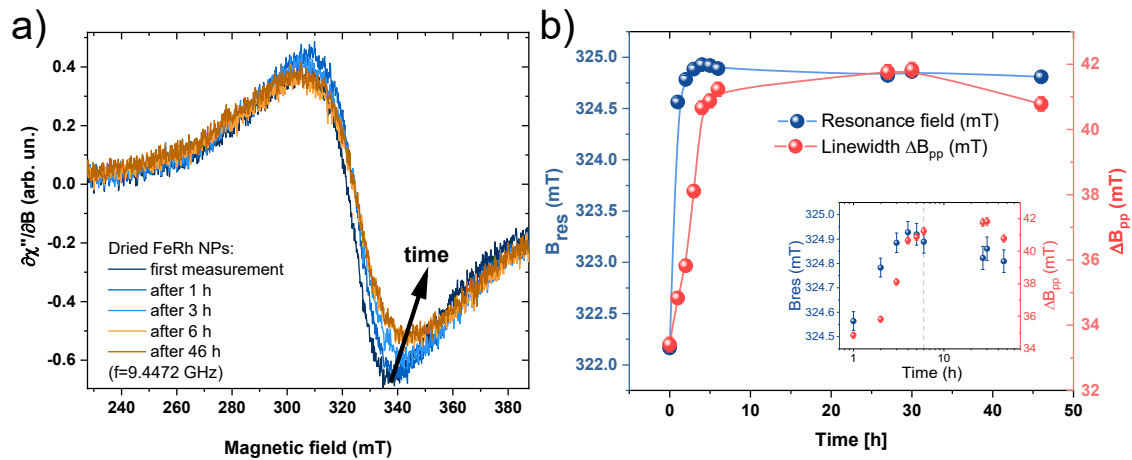

**Figure 8.** a) Evolution of FMR spectra for $Fe_{50}Rh_{50}$ nanoparticles during 46 hours after drying; b) time dependence of the corresponding resonance field and linewidth. Inset shows the data in logarithmic time scale, dot line marks 6 hours.

The evolution of the FMR signal was monitored on the dried FeRh nanoparticles exposed to a mixture of air and solvent vapour over 46 hours. During the first 6 hours, the FMR absorption spectra were recorded every hour. Few exemplary spectra are shown in Fig. 8a. Fig. 8b presents the time evolution of the resonance field and peak-to-peak linewidth. The resonance field increases with time, approaching a saturation value of 324.9±0.2 mT after 6 hours, a similar change and saturation can be seen for the linewidth, which increases from 30.8±0.5 mT for nanoparticles in solution, to 41.5±0.4 mT for dried nanoparticles within the first 10 hours after removal of acetone. This shift of the resonance field towards the isotropic value indicates a decrease of the effective anisotropy field driven by surface oxidation. Here, we like to note that due to the experimental arrangement "oxidation" does not necessarily mean due to the presence of oxygen but may also be due to an of electron transfer to chemisorbed surface ligands. The change of the resonance field can be attributed to the change of surface anisotropy, strain effects due to the lattice mismatch of oxidized and non-oxidized FeRh as well as reduction of ferromagnetic volume of the particle. The oxidation of the particles takes place within the first 6-10 hours after removal of acetone until a passivating layer is formed, similar to the one observed in $Fe_{50}Rh_{50}$ thin films [53].

As an outlook for future applications , these particles with minimized oxidation are used for nanostrand formation in a transparent polymer (figure 9) to combine the magnetic response of the and the electrical conductivity of the alloy to tune the electrical conductivity of the polymer [10, 54]. These composites could potentially be used as a window coating for electric heating while being transparent in the optical range [9]. As most of the particles are in the γ-phase, the magnetic response of the Fe$_{50}$Rh$_{50}$ nanoparticles can be shown, through the magnetic separation of the B2-phase. FeRh nano-strands are formed by applying a magnetic field (150 mT) during the formation of PMMA-FeRh composites, which is depicted in figure 9. Due to the external magnetic field, the NPs are magnetized and exhibit a local magnetic field, resulting in the attraction of the particles and alignment to strands. As most of the prepared Fe$_{50}$Rh$_{50}$ nanoparticles in the colloid are in γ-phase, which is paramagnetic at room temperature, the magnetization is too weak to form nano-strands. As the Rietveld refinement (figure 2b) shows no clear B2-phase of the sample, magnetic separation was conducted, to separate the magnetic fraction. These separated particles led to strand formation (Fig. 9b), with maximum strand lengths of 9.5 µm, which indicates a higher magnetization, comparable to ferromagnetic materials (B2-phase). An average nanostrand length of 2.6 ± 1.2 µm was observed.

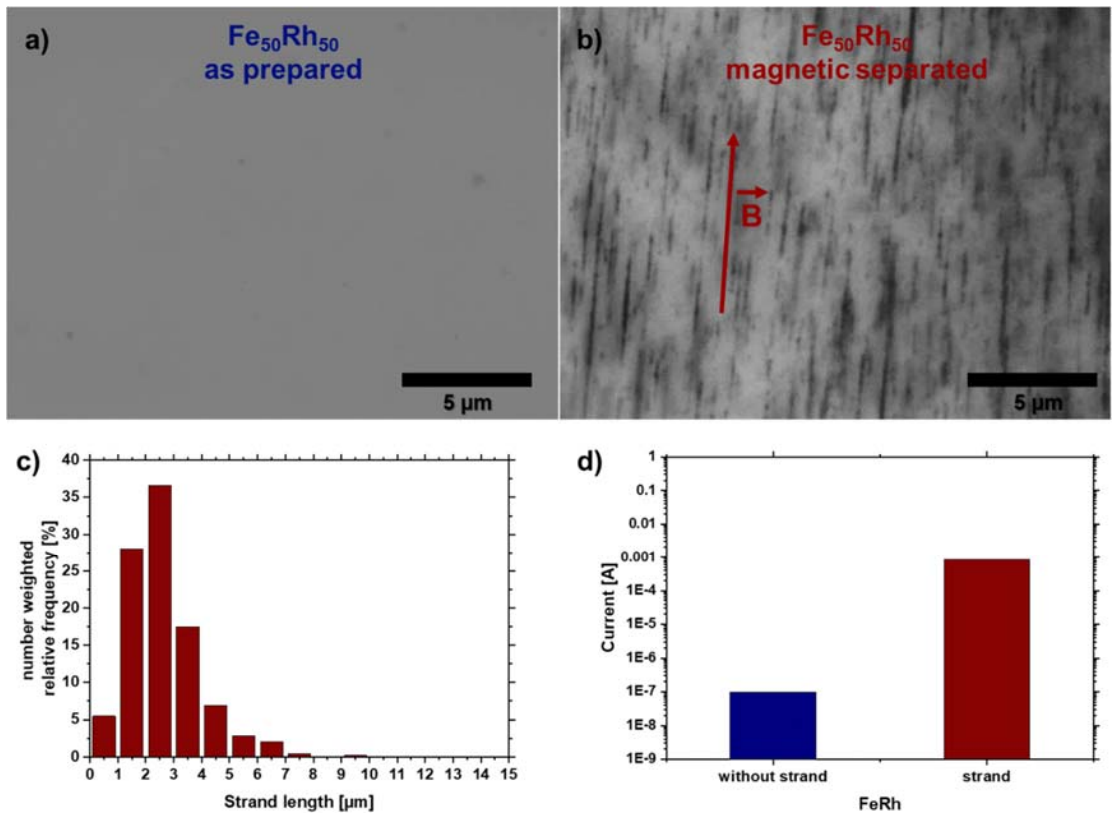

**Figure 9.** Magnetic-field induced FeRh nanostrand formation in a polymer. Light optical microscopy images of the composite containing Fe$_{50}$Rh$_{50}$ nanostrands from a) as-prepared nanoparticles and b) after magnetic separation. The red arrow denotes the direction of the magnetic field. c) Quantification of nanostrand length as extracted from microscopy through image analysis. d) Conductivity measurement of the FeRh particles without and with formed FeRh nanostrands.

The FeRh nanoparticles were further tested in terms of conductivity. Therefore, the colloid was drop casted on an interdigital electrode without the polymer. Figure 9d) shows that the electrode without strands exhibits a lower output voltage than the sample with the formed nanostrands on the electrodes. If we account for the series of resistance (shunt) of the setting, the voltage drop on this shunt yields the current. The current on the electrode without nanostrands is 95.5 nA, whereas the current of the electrode with the formed strands is 0.85 mA, which is almost a factor of 9000 higher.

**5. Conclusions**

Laser ablation in liquid (LAL) was successfully applied to synthesize metallic solid solution FeRh nanoparticles. The produced particles have monomodal size distribution with a PDI lower than 0.3 and a mean particle size of 15.24 nm for acetone and 14.46 nm for acetonitrile, respectively. Atom probe tomography confirmed the formation of solid solution FeRh nanoparticles. Three different kinds of oxygen sources were determined and analysed. It was found that residual water in the organic solvent leads to high oxidation of the particles, which was reduced by up to 50% by using a mole sieve to 24.3 at% detected oxygen in the sample. Choosing an inert or reduction gas like argon or hydrogen/nitrogen mixture can reduce the oxidation of the particles by an additional 5%. An impact of the bound oxygen in the molecule of the organic solvent could not be found. Furthermore, the oxidation behaviour after ablation in the dried state was studied by FMR, which showed a shift of the resonance field driven by surface oxidation in the first 6 hours after drying and then saturate. To apply the oxidation-minimized nanoparticles in a polymer composite, conductive nanostrands with an average nanostrand length of 2.6 ± 1.2 μm were formed from the nanoparticles after magnetic separation. The strands after magnetic separation had a factor of 9000 higher current on the electrode than single particles without the PMMA compounds. As a potential application these transparent composite could be used as a window coating for rear window defroster. Our findings show that laser ablation in liquid is a feasible method to synthesize FeRh nanoparticle with a low controllable oxidation level.

**Funding:** The authors gratefully acknowledge the funding by the German Research Foundation (DFG) within the Collaborative Research Centre / Transregio (CRC/TRR) 270 (Project-ID 405553726, projects B04, B05, B08, B09 and Z01). Bilal Gökce further thanks the DFG for funding his projects GO 2566/2-1 and GO 2566/10-1. Ayman El-zoka, Se-Ho Kim, and Baptiste Gault acknowledge financial support from the ERC-CoG-SHINE-771602.